\documentclass[prl,aps,twocolumn,superscriptaddress]{revtex4-1}

\usepackage{graphicx}
\usepackage{bm}
\usepackage{amsfonts}
\usepackage{color}
\usepackage{hyperref}

\newcommand{\bu}{\boldsymbol{u}}

\newcommand{\ocaaddress}{Universit\'e C\^ote d'Azur, CNRS, OCA,
  Laboratoire J.-L.\ Lagrange, Nice, France}

\begin{document}

\title{Dusty turbulence}

\author{J\'er\'emie Bec} \affiliation{\ocaaddress}
\author{Fran\c{c}ois Laenen} \affiliation{\ocaaddress} \author{Stefano
  Musacchio} \affiliation{Universit\'e C\^ote d'Azur, CNRS,
  Laboratoire J.-A.\ Dieudonn\'e, Nice, France}

\begin{abstract}
  The feedback forces exerted by particles suspended in a turbulent
  flow is shown to lead to a new scaling law for velocity fluctuations
  associated to a power-spectra $\propto k^{-2}$.  The mechanism at
  play relies on a direct transfer of kinetic energy to small scales
  through Kelvin--Helmholtz instabilities occurring in regions of high
  particle density contrast.  This finding is confirmed by
  two-dimensional direct numerical simulations.
\end{abstract}

\maketitle

\noindent It is common to face environmental, industrial or
astrophysical situations where impurities such as dust, droplets,
sediments, and other kinds of colloids are transported by a turbulent
fluid.  When the suspended particles have finite sizes and masses,
they detach from the flow by inertia and form uneven distributions
where intricate interactions and collisions take place.  The physical
processes at play are rather well established, leading to quantitative
predictions on the rates at which cloud droplets
coalesce~\cite{grabowski2013growth}, dust accrete to form
planets~\cite{johansen2015growth}, or heavy sediments settle in a
turbulent environment~\cite{bec2014gravity,gustavsson2014clustering}.

Still, basic and important questions remain largely open as to the
backward influence of particles on the carrier flow structure and
geometry.  Some situations involve particle mass loadings so large
that the fluid turbulent microscales are altered and, in turn, several
macroscopic processes are drastically impacted. These include spray
combustion in engines~\cite{jenny2012modeling}, aerosol saltation in
dust storms~\cite{kok2012physics}, biomixing by microorganisms in the
oceans~\cite{visser2007biomixing}, and formation of planetesimals by
streaming instabilities in circumstellar
disks~\cite{johansen2007rapid}. Currently such systems are
unsatisfactorily handled by empirical approaches or specific
treatments.  A better modelling requires identifying and understanding
the \emph{universal} physical mechanisms at play in turbulence
modulation by dispersed particles. In this spirit, we focus here on
the alteration of small scales by tiny heavy spherical particles. We
show that the fluid velocity is unstable in regions with a high
particle density contrast, leading to energy transfers shortcutting
the classical turbulent cascade. This effect leads to a novel scaling
regime of the turbulent velocity field associated to a power-law
spectrum $\propto k^{-2}$.

The fluid velocity field $\bm u$ solves the incompressible
Navier--Stokes equations: $\nabla\cdot\bu=0$ with
\begin{equation}
  \partial_t \bu + (\bu\cdot\nabla)\,\bu =
  -\frac{1}{\rho_{\rm f}}\nabla p + \nu\,\nabla^2
  \bu +\boldsymbol{f}_{\rm ext} +\boldsymbol{f}_{\rm p\rightarrow f}.
  \label{eq:navier-stokes}
\end{equation}
$\rho_{\rm f}$ is here the fluid mass density and $\nu$ its kinematic
viscosity.  A homogeneous isotropic turbulence is maintained in a
statistical steady state by an external forcing
$\boldsymbol{f}_{\rm ext}$. The fluid flow is perturbed by a
monodisperse population of small solid particles whose effects are
entailed in the force $\boldsymbol{f}_{\rm p\rightarrow f}$. These
particles are assumed sufficiently small, dilute and heavy for
approximating their distribution and dynamics in terms of fields,
namely a mass density $\rho_{\rm p}$ and a particle velocity field
$\boldsymbol{v}_{\rm p}$ satisfying
\begin{eqnarray}
  && \partial_t \rho_{\rm p} + \nabla\cdot(\rho_{\rm p}\boldsymbol{v}_{\rm
     p}) = 0 \label{eq:rhop} \\
  && \partial_t \boldsymbol{v}_{\rm p} + (\boldsymbol{v}_{\rm
     p}\cdot\nabla)\,\boldsymbol{v}_{\rm p} = -\frac{1}{\tau_p} \left(
     \boldsymbol{v}_{\rm p}-\boldsymbol{u}\right), \label{eq:vp}
\end{eqnarray}
where $\tau_{\rm p} = 2\rho_{\rm s}\,a^2/(9\rho_{\rm f}\,\nu)$ is the
particles response time, $a$ being their radius and $\rho_{\rm s}$ the
mass density of the material constituting the particles.  The
hydrodynamical system~(\ref{eq:rhop})-(\ref{eq:vp}) has proven to be a
valid approximation for relatively small Stokes numbers
$St = \tau_{\rm p} / \tau_{\rm f}$~\cite{boffetta2007eulerian}, that
is when the particle response time is smaller than the smallest active
timescale $\tau_\mathrm{f}$ of the fluid flow.  In this limit, fold
caustics appear with an exponentially small
probability~\cite{wilkinson2006caustic,gustavsson2011distribution},
preventing the development of multivalued branches in the particle
velocity profile and thus ensuring the validity of a hydrodynamical
description.

The force exerted by the particles on the fluid reads
\begin{equation}
  \boldsymbol{f}_{\rm
    p\rightarrow f} = \frac{1}{\tau_{\rm p}}\,\frac{\rho_{\rm p}}{\rho_{\rm f}}
  \left(\boldsymbol{v}_{\rm p}-\boldsymbol{u}\right).
  \label{eq:fpf}
\end{equation}
It is proportional to the mass density of the dispersed phase and thus
combines the heaviness of the particles with their number density. The
strength of feedback is measured by the comprehensive non-dimensional
parameter $\Phi = \langle\rho_{\rm p}\rangle / \rho_{\rm f}$.  It
involves the particle density spatial average
$\langle\rho_{\rm p}\rangle = N_{\rm p}\,m_{\rm p}/\mathcal{V}$, where
$N_{\rm p}$ is the total number of particles, $m_{\rm p}$ their
individual mass, and $\mathcal{V}$ the volume of the domain. All these
quantities being conserved by the dynamics, so is the coupling
parameter $\Phi$.

We first draw some straightforward comments pertaining to the limit of
small Stokes numbers.  There, particles almost follow the flow with a
tiny compressible correction~\cite{maxey1987gravitational}, namely
$\boldsymbol{v}_{\rm p} \approx \boldsymbol{u} - \tau_{\rm
  p}\,\boldsymbol{a}$,
where
$\boldsymbol{a} = \partial_t \boldsymbol{u} +
(\boldsymbol{u}\cdot\nabla)\boldsymbol{u}$
denotes the fluid flow acceleration field.  The feedback force exerted
on the fluid is hence, to leading order,
\begin{equation}
  \boldsymbol{f}_{\rm
    p\rightarrow f} (\boldsymbol{x},t) \approx - \frac{1}{\rho_{\rm
      f}}\,\rho_{\rm
    p}(\boldsymbol{x},t)\,\boldsymbol{a}(\boldsymbol{x},t)
  \label{eq:fpf_smlst}
\end{equation}
The effect of particles can thus be seen as an \emph{added mass},
which does not depend upon their response time and is responsible for
an increase of the fluid inertia. The fluid is accelerated as if it
has an added density equal to that of the particles. Such
considerations predict that the presence of particles decreases the
effective kinematic viscosity of the fluid and thus increases its
level of turbulence. This vision is however too naive as it overlooks
the spatial fluctuations of the particle density. It is indeed known
that an even infinitesimal inertia of the particles creates extremely
violent gradients of their density through the mechanism of
preferential concentration. As we will now see, these variations are
responsible for instabilities that shortcut the turbulent energy
cascade by directly transferring kinetic energy to the smallest
turbulent scales.

\begin{figure}[h]
  \begin{center}
    \includegraphics[width=\columnwidth]{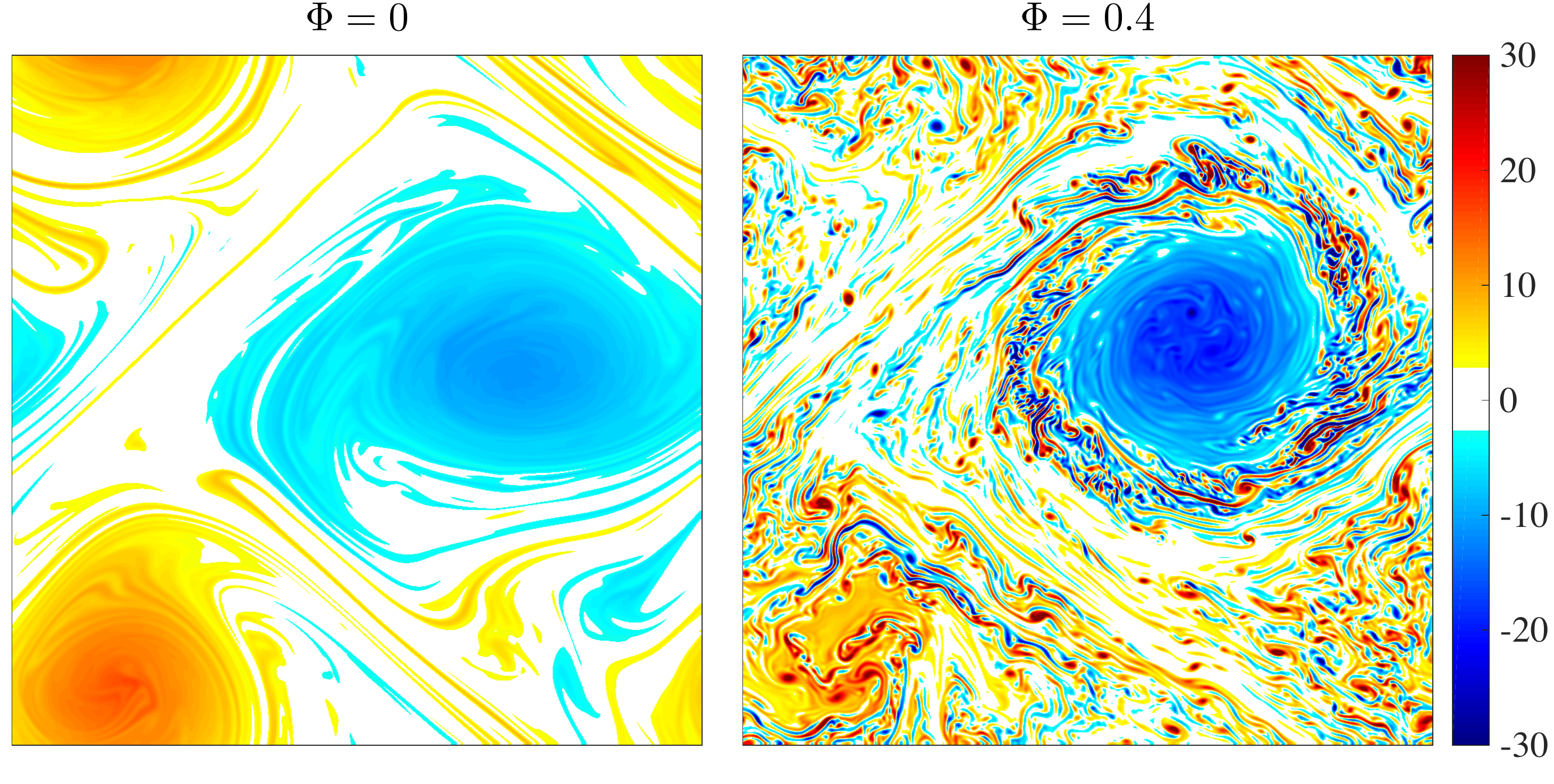}\\[0pt]
    \includegraphics[width=\columnwidth]{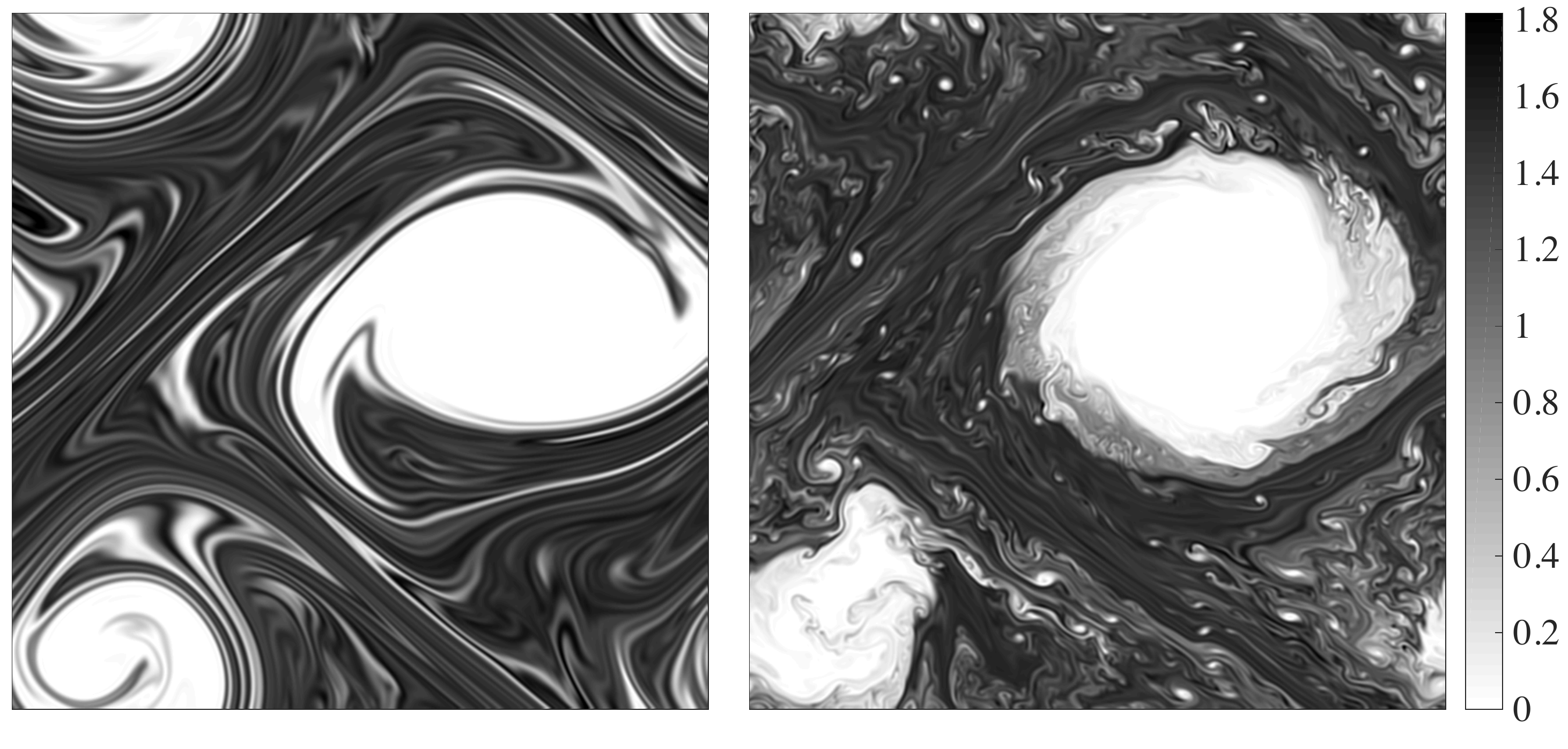}
    \caption{\label{fig:snap_vort_rho} (color online) Upper frames:
      Snapshots of the (scalar) fluid vorticity
      $\boldsymbol{\omega} = \nabla \times\boldsymbol{u}$ in two
      dimensions for both the passive case with no feedback
      ($\Phi=\langle\rho_{\rm p}\rangle / \rho_{\rm f}=0$, left) and
      when the particles exert a back reaction on the flow
      ($\Phi=0.4$, right). Lower frames: normalized particle density
      $\rho_{\rm p} / \langle\rho_{\rm p}\rangle$ at the same instants
      of time. The two cases (without and with coupling) correspond to
      different realizations of the external forcing; the $x$ and $y$
      axis were however shifted in order to locate large-scale
      structures at approximately the same position.}
    \label{fig:snaps}
  \end{center}
\end{figure}

A key attribute of turbulence is the vigorous local spinning of the
fluid flow, weighed by the vorticity
$\boldsymbol{\omega} = \nabla \times\boldsymbol{u}$.  The effect of
particles on the vorticity dynamics is entailed in the curl of the
feedback force (\ref{eq:fpf}), reading
\begin{equation}
  \nabla\times \boldsymbol{f}_{\rm
    p\rightarrow f} =  \frac{1}{\rho_{\rm f}\,\tau_{\rm
      p}}\left[ \rho_{\rm
      p}\left(\boldsymbol{\omega}_{\rm
        p}-\boldsymbol{\omega}\right)  +\nabla\rho_{\rm p} \times
    \left(\boldsymbol{v}_{\rm p}-\boldsymbol{u}\right)\right],
\end{equation}
where
$\boldsymbol{\omega}_{\rm p} = \nabla \times\boldsymbol{v}_{\rm p}$ is
the vorticity of the dispersed phase.  The action of particles is thus
twofold. The first term accounts for a friction of the fluid vorticity
with that of the particles, which amounts at small Stokes numbers to
the above-mentioned added-mass effect.  The second term gives a source
of vorticity proportional to the gradients of particle density. The
combined effects of preferential concentration and turbulent mixing is
responsible for very sharp spatial variations of $\rho_{\rm p}$.
Centrifugal forces indeed eject heavy particles from coherent vortical
structures~\cite{maxey1987motion} and Lagrangian transport stretches
particles patches in stirring
regions~\cite{haller2000lagrangian}. This leads to the development of
substantial fluctuations of $\nabla\rho_{\rm p}$, as illustrated in
two dimensions on the left panels of Fig.~\ref{fig:snaps}. This
mechanism creates regions with very strong shear in the fluid flow,
which, in turn, develop small-scale vortical structures through
Kelvin--Helmholtz instability. It is indeed well known that flows
presenting a quasi-discontinuity of velocity are linearly unstable and
develop wavy vortical streaks at the interface of the two motions
(see, \emph{e.g.},~\cite{chandrasekhar1961hydrodynamic}).  Such
phenomenological arguments thus suggest that the feedback of particles
lead to the formation of small-scale eddies, as can be seen in the
right panels of Fig.~\ref{fig:snaps}. Particles thus actively
participate in the transfer of kinetic energy toward the smallest
turbulent scales.

This effect and the resulting modification of the fluid flow scaling
properties can be quantified by examining the scale-by-scale kinetic
energy budget given by K\'arm\'an--Howarth--Monin relation (see,
\emph{e.g.},~\cite{frisch1995turbulence}).  Denoting the velocity
increment over a separation $\boldsymbol{r}$ by
$\delta_r \boldsymbol{u} = \boldsymbol{u}'-\boldsymbol{u}$ with
$\boldsymbol{u}' = \boldsymbol{u}(\boldsymbol{x}+\boldsymbol{r},t)$
and $\boldsymbol{u} = \boldsymbol{u}(\boldsymbol{x},t)$, one can
easily check that statistically homogeneous solutions to the
Navier--Stokes equation (\ref{eq:navier-stokes}) satisfy
\begin{eqnarray}
  \frac{1}{2}\partial_t\left\langle \boldsymbol{u}\cdot \boldsymbol{u}'
  \right\rangle &=& \frac{1}{4} \nabla_r\cdot \left\langle \left|
                    \delta_r \boldsymbol{u} \right|^2 \delta_r \boldsymbol{u}
                    \right\rangle + \nu\,\nabla_r^2\left\langle \boldsymbol{u}\cdot
                    \boldsymbol{u}' \right\rangle \nonumber \\
                && +\left\langle \boldsymbol{u}\cdot
                   \bar{\boldsymbol{f}}_{\rm ext} \right\rangle
                   +\left\langle \boldsymbol{f}_{\rm p\to f}\cdot
                   \bar{\boldsymbol{u}} \right\rangle,
                   \label{eq:khm}
\end{eqnarray}
where the overbar denotes the average over the two points located at
$\boldsymbol{x}\pm\boldsymbol{r}$, that is
$\bar{\boldsymbol{f}} =
[\boldsymbol{f}(\boldsymbol{x}+\boldsymbol{r},t)
+\boldsymbol{f}(\boldsymbol{x}-\boldsymbol{r},t)]/2$.
In classical stationary turbulence, the above relation suggests a
balance between the non-linear transfer term (first term on the
right-hand side) and viscous dissipation (second term), leading for
isotropic flows to the celebrated Kolmogorov $4/5$ law.  In the
presence of coupling with particles, this equilibrium is broken by the
feedback force. In the asymptotics $St \ll 1$ of low inertia, this
force is approximated by~(\ref{eq:fpf_smlst}), so that its
contribution to (\ref{eq:khm}) reads
\begin{equation}
  \left\langle \boldsymbol{f}_{\rm p\to f}\cdot\bar{\boldsymbol{u}}
  \right\rangle \approx \frac{1}{\rho_{\rm f}} \left\langle
    \rho_{\rm p}\, \boldsymbol{a} \cdot \boldsymbol{u} \right\rangle
  - \frac{1}{\rho_{\rm f}} \left\langle \rho_{\rm p}\, \boldsymbol{a} \cdot
    \delta_r\boldsymbol{u} \right\rangle.
  \label{eq:contrib_fpf}
\end{equation}
The first term on the right-hand side involves the correlation between
the particle density field and the instantaneous power acting on fluid
elements. To leading order when $St\to0$, we have
$\left\langle \rho_{\rm p}\, \boldsymbol{a} \cdot \boldsymbol{u}
\right\rangle \approx \left\langle\rho_{\rm p}\right\rangle
\left\langle\boldsymbol{a} \cdot \boldsymbol{u} \right\rangle = 0$.
Non-vanishing corrections at small but finite Stokes numbers might
arise from a combined effect of the small compressibility of the
particle velocity together with the biased sampling due to
preferential concentration, as already seen for the radial
distribution
function~\cite{balkovsky2001intermittent,chun2005clustering}.  However
such correlations are in the best case of the order of $St^2$.  The
second term on the right-hand side of (\ref{eq:contrib_fpf}) does not
vanish in the limit $St\to0$ and thus gives the dominant
contribution. 

Such arguments lead to predict that the scale-by-scale energy
balance~(\ref{eq:khm}) reduces in the inertial range to
\begin{equation}
  \frac{1}{4} \nabla_r\cdot \left\langle \left|
      \delta_r \boldsymbol{u} \right|^2 \delta_r \boldsymbol{u}
  \right\rangle  \simeq \frac{1}{\rho_{\rm f}} \left\langle \rho_{\rm
      p}\, \boldsymbol{a} \cdot \delta_r\boldsymbol{u} \right\rangle.
  \label{eq:khm_red}
\end{equation}
Now, assuming that the fluid velocity field obeys some scaling
property $\delta_r \boldsymbol{u} \sim r^h$, one deduces from the
above balance that $3h-1=h$, and thus $h =1/2$.  Such a scaling
behavior is associated to an angle-averaged kinetic energy power
spectrum $E(k) \propto k^{-2}$.

In order to test such prediction, we perform two-dimensional
simulations of the fluid-particle system defined by
(\ref{eq:navier-stokes}), (\ref{eq:rhop}), (\ref{eq:vp}), and
(\ref{eq:fpf}) in a periodic domain.  We make use of a
Fourier-spectral solver with $1024^2$ collocation points for
estimating spatial derivatives and of a second-order Runge-Kutta
scheme for time marching.  We focus on the direct enstrophy cascade,
so that the external forcing $\boldsymbol{f}_{\rm ext}$ is the sum of
an Ekman friction with timescale $1/\alpha$ and of a random Gaussian
field $\eta$ white noise in time and concentrated at wavenumbers
$|\boldsymbol{k}|\le 2$.  We make use of hyper-viscosity and
hyper-diffusivity (fourth power of the Laplacian) in order to maximize
the extent of the inertial range and prevent Eqs.~(\ref{eq:rhop}) and
(\ref{eq:vp}) from blowing up. The particle response time is fixed in
such a way that
$St = \tau_{\rm p}\,\langle \omega^2\rangle^{1/2} \approx 10^{-2}$ in
the uncoupled case and various values of the coupling parameter
$\Phi=0$, $0.1$, $0.2$, and $0.4$ are simulated.

Figure~\ref{fig:snaps} shows snapshots of the fluid vorticity field
together with the particle density field, without and with coupling
between the two phases. In the absence of feedback from the particles
(left panels), the flow develops the traditional picture of
two-dimensional direct cascade consisting of large-scale vortices
separated by a bath of filamentary structures where enstrophy is
dissipated. The particles density field is characterized by large
voids in the vortical structures separated by a filamentary
distribution that is symptomatic of turbulent mixing. These
qualitative pictures are strongly altered when the particle feedback
is turned on. In the presence of coupling (right panels), the fluid
flow still shows large-scale structures but which are this time
surrounded by a bath of small-scale vortices. These eddies form wavy
structures along the lines associated to quasi-discontinuities of the
particle density field. This is a clear signature that
Kelvin--Helmholtz instability is at play.

\begin{figure}[h]
  \begin{center}
    \includegraphics[width=\columnwidth]{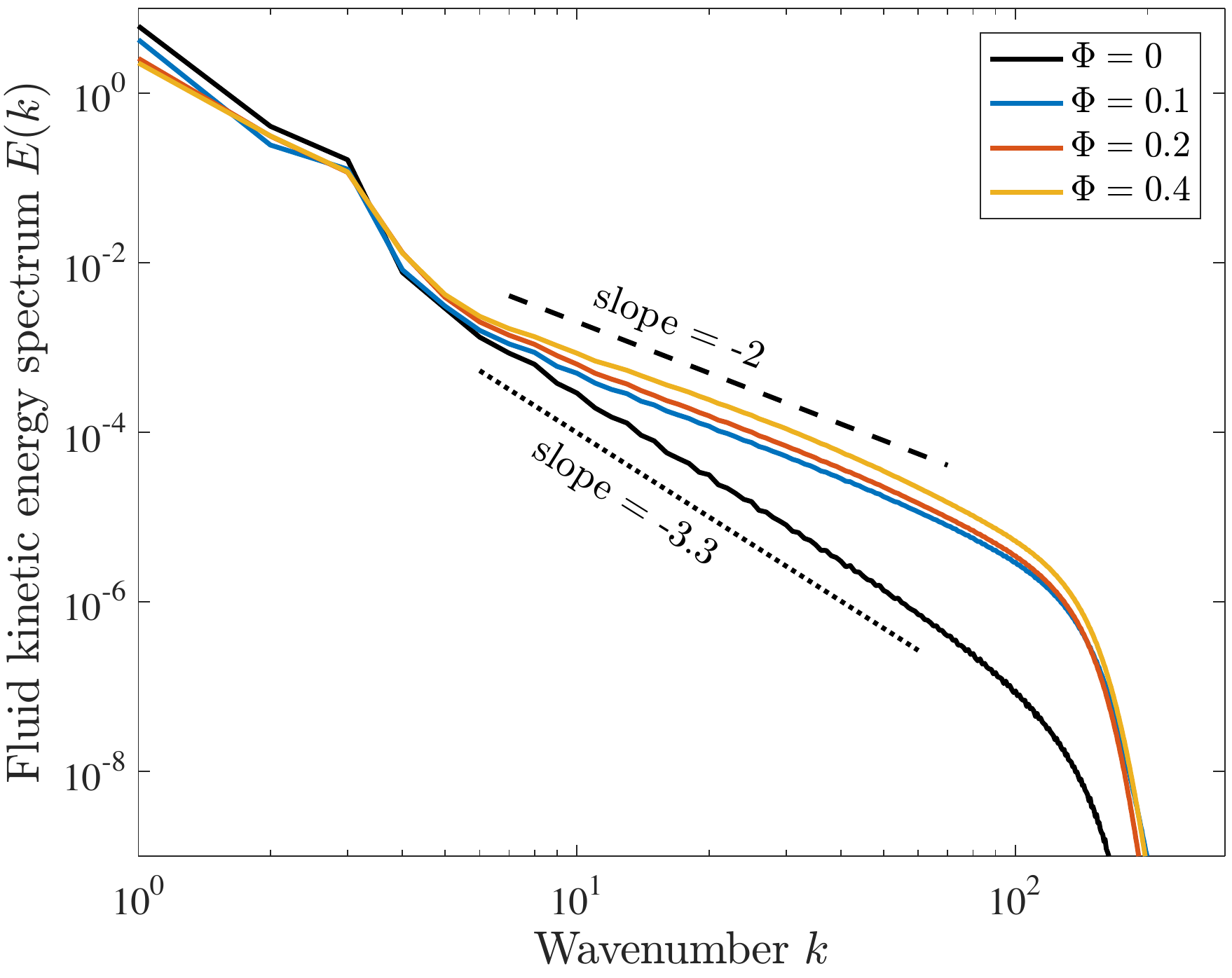}
    \caption{\label{fig:EnergySpectra} (color online) Angle-averaged
      kinetic energy power spectra of the fluid velocity represented
      for various values of the coupling parameter $\Phi$, as
      labelled.}
    \label{fig:energyspectra}
  \end{center}
\end{figure}
Figure~\ref{fig:energyspectra} shows the angle-averaged power spectra
of the fluid kinetic energy obtained when varying the coupling
parameter. In the case of no feedback ($\Phi=0$), the specific choices
of the Ekman coefficient $\alpha$ and of the energy injection
amplitude yield a kinetic energy spectrum $E(k)\propto k^{-\delta}$
with $\delta \approx 3.3$.  For any non-vanishing value of the
coupling parameter $\Phi$, one observes remarkable changes in the
spectral behavior of the fluid velocity. The first effect is a clear
decrease of the total kinetic energy. Similarly to what is obtained in
the asymptotic of large Stokes numbers~\cite{laenen2017modulation},
this is due to a net dissipative effect of the coupling with the
particle phase. However this impacts only the largest scales of the
flow and the smaller scales experience an increase in their energy
content. The inertial-range is characterized by a shallower power
spectrum with an exponent close to $-2$, as expected from above
arguments. Dissipative scales are shifted toward larger wavenumbers,
as a consequence of the added-mass effect induced by particles which
decreases the effective kinematic viscosity of the fluid loaded by
particles.

\begin{figure}[h]
  \begin{center}
    \includegraphics[width=\columnwidth]{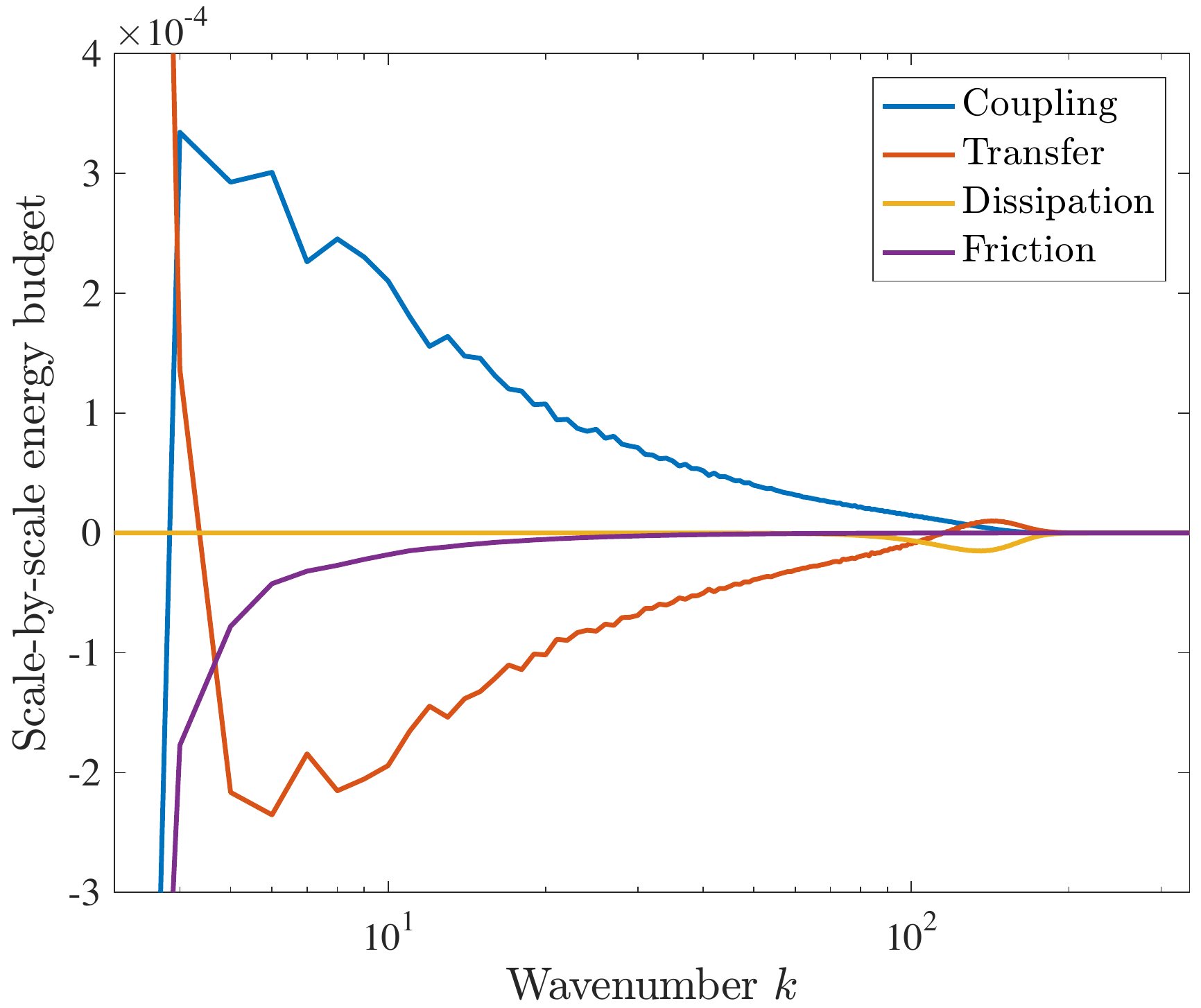}
    \caption{\label{fig:EnergyBudget} (color online) Angle-averaged
      Fourier amplitudes of the various terms contributing to the
      kinetic energy budget~(\ref{eq:khm}) shown here for
      $\Phi=0.4$. Coupling stands from the contribution of the forces
      exerted by the particles on the fluid, transfer for the
      nonlinear advection terms, dissipation for viscous forces and
      friction for Ekman damping.}
    \end{center}
\end{figure}
Further insight is given by measuring the amplitude of the various
terms entering in the energy
budget~(\ref{eq:khm}). Figure~\ref{fig:EnergyBudget} shows the
angle-averaged amplitude of their Fourier transforms with respect to
the separation $\boldsymbol{r}$. One observes that the non-linear
transfer term gives a positive contribution at small wavenumbers. This
is a strong signature of two-dimensional turbulence for which,
conversely to three dimensions, the nonlinear terms are not
transferring kinetic energy toward small scales participating to its
accumulation at largest lengthscales of the flow.  This term is
exactly compensated by the linear Ekman friction and the coupling with
the particle phase which are both negative and of the same
order. Coupling is thus pumping energy at large scales but restitutes
it at larger wavenumbers as it is positive for $k\ge4$. In the
inertial range for $10\lesssim k\lesssim 100$ where both the
contribution of Ekman friction and viscous dissipation are negligible,
it is exactly compensated by a negative value of the nonlinear
transfer term. Both curves decrease as $k^{-1}$, in agreement with the
scaling observed earlier. At the smallest scales, coupling becomes
negligible, nonlinear transfer changes sign and is compensated by
viscous dissipation. The whole two-dimensional picture thus confirms
the prediction made above.

We have thus evidenced from this work a new regime of turbulent flow
where the feedback of suspended particles onto the fluid flow
dominates inertial-range energy transfers.  This regime is evidenced
by numerical simulations in two dimensions but such strong effects
should also be present in three dimensions, at least at sufficiently
small scales.  A remarkable feature of this turbulent enhancement due
to dust-like particles is the creation of small-scale eddies whose
spectral signature is a $k^{-2}$ power-law range for the fluid
velocity. These vortices profoundly affect particle concentration. On
the one-hand, their spatial distribution tends to weaken large-scale
inhomogeneities, to reduce potential barriers to transport and enhance
mixing. On the other hand, the dispersion in the flow and the
interactions between these long-living structures trigger density
fluctuations that are much more intense than in the absence of
coupling between the two phases.  Such effects clearly need being
investigated in a more systematic manner: They might indeed strongly
modify at both qualitative and quantitative levels the rate at which
particles interact together.

We acknowledge useful discussions with G.~Krstulovic.  The research
leading to these results has received funding from the French Agence
Nationale de la Recherche (Programme Blanc ANR-12-BS09-011-04).

\bibliography{biblio}

\end{document}